\documentclass[a4paper,12pt]{article}
\pdfoutput=1 

\usepackage{jheppub} 
\usepackage{hyperref}
\hypersetup{
    bookmarks=true,         
    unicode=false,          
    pdftoolbar=true,        
    pdfmenubar=true,        
    pdffitwindow=false,     
    pdfstartview={FitH},    
    pdftitle={My title},    
    pdfauthor={Author},     
    pdfsubject={Subject},   
    pdfcreator={Creator},   
    pdfproducer={Producer}, 
    pdfkeywords={keyword1} {key2} {key3}, 
    pdfnewwindow=true,      
    colorlinks=true,       
    linkcolor=red,          
    citecolor=purple,        
    filecolor=magenta,      
    urlcolor=blue,           
    linktocpage=true
}
\usepackage{graphics}
\usepackage{rotating}
\usepackage{subfigure}
\usepackage{pstricks}
\usepackage{color}
\usepackage{amsfonts}
\usepackage{mathrsfs,amsmath}
\usepackage{epsfig}
\usepackage{amsmath,amssymb,amsthm,graphicx,latexsym}
\usepackage{ulem}
\usepackage{multirow}

\newcommand{\be}{\begin{equation}}
\newcommand{\ee}{\end{equation}}
\newcommand{\bea}{\begin{eqnarray}}
\newcommand{\eea}{\end{eqnarray}}
\newcommand{\bml}{\begin{subequations}}
\newcommand{\eml}{\end{subequations}}
\newcommand{\bfig}{\begin{figure}}
\newcommand{\efig}{\end{figure}}

\newcommand{\del}{\delta}

\newcommand{\slashed}{\hspace{-1.1ex}/}

\newcommand{\bmat}{\begin{pmatrix}}
\newcommand{\emat}{\end{pmatrix}}
\begin{document}
	\title{\textsc{\fontsize{25}{17}\selectfont \sffamily \bfseries \textcolor{purple}{EFT of Inflation: Reflections on CMB and Forecasts on LSS Surveys}}}
	
	\author[a]{Abhishek Naskar,}	
	\author[b,c]{Sayantan Choudhury,}
	 \author[d]{Argha Banerjee,}		
	\author[a,e]{Supratik Pal}
	\affiliation[a]{Physics and Applied Mathematics Unit, Indian Statistical Institute, 203 B.T. Road,
	Kolkata  700108, India}
   \affiliation[b]{School of Physical Sciences,  National Institute of Science Education and Research, Bhubaneswar, Odisha - 752050, India}
   \affiliation[c]{Homi Bhabha National Institute, Training School Complex, Anushakti Nagar, Mumbai - 400085, India}
	\affiliation[d]{School of Physics and Astronomy, University of
Minnesota, Minneapolis 55455, USA
	}
	\affiliation[e]{Technology Innovation Hub on Data Science, Big Data Analytics and Data Curation, 
	Indian Statistical Institute, 203 B.T. Road,
	Kolkata  700108, India}
	\emailAdd{abhiatrkmrc@gmail.com, sayantan.choudhury@niser.ac.in, baner124@umn.edu, supratik@isical.ac.in}

	\abstract{We investigate the possibility of constraining parameters of Effective Field Theory 
	(EFT)
	of inflation with upcoming Large Scale Structure (LSS) surveys in order to have a better understanding of inflationary 
	dynamics. With the  development of the construction algorithm of EFT, we arrive at a properly truncated action 
	for the entire scenario.  Using this, we compute the two-point correlation function for quantum fluctuations from Goldstone modes
	and related inflationary observables in terms of coefficients of relevant EFT operators.
 	We then perform Fisher matrix forecast analysis to estimate the possible error bars on the parameters of
	EFT as well as on the inflationary parameters using two upcoming LSS surveys, namely, LSST and EUCLID.
	}


	\maketitle
	\flushbottom
	\section{Introduction}
 The great success of inflationary paradigm lies in its indigenous ability to explain the origin of 
 Large Scale Structure (LSS) that is observable today. Perturbations generated during inflation
 first exit the horizon and get frozen, and then re-enter horizon and grows due to
  gravitational instability to give rise to LSS.
The usual method to study inflationary cosmology is to start with a scalar field characterized by a potential and analyze the effects of  quantum fluctuations
 of the field. An extensive list of the potentials that are usually considered is given in \cite{martin} and an analysis of the models while comparing with CMB data can be found in \cite{ring}. In recent years an alternative approach  to describe perturbations based on EFT has been introduced \cite {crem, weinberg}.
Here, considering that the inflaton field $\phi$ is a scalar under all diffeomorphisms (diffs) but the fluctuation $\delta \phi$ is  scalar only under spatial diffs and transforms under the time diffs., one can construct an  effective action in unitary gauge where fluctuation in inflaton is zero, by allowing every possible gravitational fluctuation interaction terms that respect the symmetry. Using  {\it St$\ddot{u}$ckelberg mechanism} \cite{stue1, stue2} the gravitational perturbation can be written in terms of massless Goldstone boson and curvature perturbation can also be related to the Goldstone boson. One interesting feature of this scenario is that at high energies the Goldstone boson decouples from gravitational fluctuation and the corresponding Lagrangian becomes very simple to study perturbation and in different parameter space of EFT coefficients one can retrieve different inflationary models. Also
the EFT parameters characterize the primordial observables.
	Further developments of EFT of inflation from different prescriptions have also been done in	\cite{senatore,yamaguchi,zal,palma,gong}. 
In \cite{senatore2,crem2}	 it has been shown how  one 
can  generate large non-Gaussianity from single field inflation with sound speed $c_s<1$  using EFT.  
Some recent works on diverse aspects of EFT of inflation can also be found in \cite{cai,gorbenko,cai2,gong2,giblin,rostami,motohashi,bordin,Bartolo:2010bj,Bartolo:2010di,Choudhury:2017glj,Naskar:2018rmu,Naskar:2019shl,Naskar:2020vkd}. 
	
	As inflationary perturbations seed the LSS, physics of 
     LSS  provides us with a plethora of information about the initial condition and subsequent evolution of the universe.	
    Theoretical studies of LSS are usually  materialised by a linear perturbation theory (LPT) (see, for example, \cite{colombi}) keeping in mind the observational fact that
    on the large scale the perturbations are linear. In this regime the Fourier modes of perturbations evolve independently of each other keeping the stochastic properties
	of the primordial fluctuations intact. However, at small scales the fluctuations become non-linear and the modes no longer remain 
	independent of each other. Consequently, small scales affect the evolution of large scales \cite{goroff, scocci2}.   
	Studies of small scale physics thus become a bit tricky due to the overlapping of modes. 
	Additionally, there are several ambiguities at small scales that leads to degeneracies among models, and a fairly clear understanding of non-linear regime 
	from a non-linear cosmological perturbation theory (NLCPT) is 
	yet to be achieved.	 
	Early attempts in this direction were made 
	by  calculating the next-to-leading order correction  for Gaussian initial condition   \cite{scocci3, suto, jain}. In this papers, for the linear regime,  the usual Harrison-Zel'dovich power spectrum
	with  power law behaviour $P(k)\propto k^n$ have been considered for $k_{IR}<k<k_{UV}$; where $k_{IR}$ and $k_{UV}$ are, respectively, infrared and ultra-violet cutoffs for the power spectrum;
	 whereas for the non-linear regime,  the exponent $n$ of the power spectrum was 
	constrained by calculating the loop contributions and subjecting them to N-body simulation.  Subsequently,  in \cite{scocci}  the authors introduced renormalized perturbation 
	to NLCPT and tried to cancel the cutoff dependence of the loops in small scales. 
	
	Given the difficulties in proposing a proper NLCPT, searches for alternative ideas of explaining the non-linear regime are in vogue.  One such
	 useful  alternate technique is based on 
	Effective Field Theory (EFT henceforth) approach. In   \cite{carrasco} the authors describe how one can conveniently develop a perturbation theory based on
	EFT that can, to a considerably good extent, explain 
	 LSS perturbations. Here the small scales perturbations were integrated out beyond a UV cutoff and their effects on the large scales were studied.	
	 	Further developments of issues like renormalization, higher loop calculation, bispectrum calculation etc. can be found in \cite{pajer, green, baldauf,senatore3}.  
		The idea of \cite{carrasco} has also been
	extended to non-Gaussian initial condition in \cite{baumann} for a nearly scale invariant power spectrum. 
\cite{lewandowski,perko,lewandowski2,bella,munshi,Hertzberg:2012qn} also discuss some very recent developments 
in the direction of LSS from EFT. In \cite{piet,pelo1} the authors pursue an alternative but related technique to \cite{carrasco} 
called Coarse Grained Perturbation Theory where the UV source terms themselves are  measured using N-body simulations.
			
From observational point of view LSS surveys can play a crucial role in probing primordial observables 
with great precision. Probing early universe using LSS surveys has been explored to some extent in
 \cite{Agarwal:2013qta,DiDio:2013sea,LHuillier:2017lgm,Chandra:2018vkd}.
As is well-known,  LSS surveys usually have access to wider range of scales to analyze  than 
the CMB observations. For this reason,  they  have great potential to further constrain primordial parameters with greater accuracy. 
In the same vein, it is quite expected that upcoming LSS surveys like LSST \cite{Zhan:2017uwu,Abell:2009aa}, 
EUCLID \cite{Laureijs:2011gra,Amendola:2016saw}, DESY \cite{Aghamousa:2016zmz} can improve the constraints of 
primordial observables significantly. 
In this article we would  like to take up  a rigorous approach to address this interesting and timely issue.
To elaborate our scheme little bit, we would 
 go through a two-step process, namely, 
\begin{enumerate}
\item[(i)] to construct a consistent EFT of inflation and to obtain the observable parameters in the light of CMB therefrom,

\item [(ii)] to forecast on  the set of parameters (EFT parameters as well as the parameters related to inflation) for couple of upcoming LSS surveys.
\end{enumerate}

To this end, we would construct primordial scalar curvature powerspectrum staring 
from the framework of EFT of inflation, that will take into account the effects of   EFT parameters. We would then make use of Fisher matrix forecast analysis and  will  forecast on  
  these parameters for 
upcoming LSS surveys of LSST and EUCLID. For forecast, we will choose different fiducial values of the parameters from the already existing constraints on them as given by  Planck 2018 data \cite{Akrami:2018odb}. This will not only
make the analysis consistent with latest CMB observations but also  help us make a comparison among results obtained from different fiducial values.
This study has two fold importance: $(i)$ we can have an idea
about the EFT parameter space from upcoming surveys \textit{i.e} we will have an idea about preferred 
type of inflationary models
 by upcoming data, $(ii)$
and  this will give us an idea about how well can
these upcoming LSS surveys extract  primordial
physics without considering a particular model of inflation as such.

The plan of the paper is as follows. In Sec \ref{sec-2} we review the framework of EFT of inflation briefly, in Sec \ref{sec-3} the primordial scalar powerspectrum from EFT framework is calculated, in Sec \ref{sec-4} 
the perturbative approach of LSS dynamics is reviewed briefly and 
in Sec \ref{sec-5}  Fisher forecast method is used 
estimate the EFT parameters that characterize the primordial powerspectrum. We end up with a  summary of out results and a brief discussion.

		
\section{Construction of consistent EFT for inflation}	\label{sec-2}

Let us begin with the construction of  a consistent EFT which is applicable to both inflation and LSS. Although our construction is somewhat motivated by
some earlier works in this field \cite{senatore,yamaguchi,zal,palma,gong,crem,weinberg, cai,gorbenko,cai2,gong2,giblin,rostami,motohashi,bordin,piazza,vernizzi},
we develop the framework in our own language that will also help in arriving at  a consistent description of inflationary framework as well as large scale structures.

\subsection{Truncated action for EFT}\label{2a}
	
The construction algorithm of the EFT action  is as follows: 
\begin{enumerate}
\item In this description we will deal with quasi de Sitter background with spatial slicing, which can be identified as a single physical clock using which one can describe the cosmological evolution smoothly. This can conveniently be  described by  time evolution of a single scalar field $\phi(t)$. In order  to deal with the cosmological perturbations in this framework one needs to choose unitary gauge where all the slicing coincides with hypersurfaces of constant time i.e.
$\delta\phi(t,{\bf x})=0$.
This guarantees that cosmological perturbations are fully described by metric fluctuation alone and no other explicit contribution from the perturbation of the scalar field appear in the theory. 

\item The temporal diffeomorphisms are completely fixed by this specific gauge choice. Consequently, the metric fluctuation from graviton is described by three degrees of freedom: two projections of  tensor helicities and one scalar mode. One can further interpret that the scalar perturbation is completely eaten by metric, 
which mimics the role of Goldstone boson as appearing in the context of any $SU(N)$ non abelian gauge theory. 

\item In order  to construct a  general, model independent action of EFT in this framework, one needs to define a class of relevant operators $\delta{\cal O}_{\rm \bf EFT}$ that are function of the background metric and invariant under time dependent spatial diffeomorphisms in linear regime as described by following coordinate transformation
\be x^{i}\rightarrow x^{i}+\xi^{i}(t,{\bf x}) ~~~\forall~~~ i=1,2,3.\ee 
This implies that the spatial diffeomorphism is broken by an amount $\xi^{i}(t,{\bf x})~\forall~i=1,2,3$. For example, we consider temporal part of the background metric $g^{00}$ and the extrinsic curvature of hypersurface with constant time extrinsic curvature $K_{\mu\nu}$, which are transforming like a scalar and tensor under time dependent spatial diffeomorphisms in linear regime respectively. It is important to mention here that, in the present context the preferred slicing of the spacetime is described by a function $\tilde{t}(x)=\tilde{t}(t,{\bf x})$, which has a timelike gradient, $\partial_{\mu}\tilde{t}(x)=\delta^{0}_{\mu}$. Such a function is necessary to implement time diffeomorphism in the linear regime. Covariant derivatives of $\partial_{\mu}\tilde{t}(x)$ (essentially the covariant derivatives of unit normal vector $n_{\mu}$  defined later) and its projection tensor $K_{\mu\nu}$ play significant role in the construction of the EFT action in unitary gauge. In the unitary 
gauge the 
temporal coordinate coincide with this specified function. This restricts all other additional degrees of freedom appearing in this function and the final form of the EFT action is free from any ambiguity arising from such contributions. 

\item It is worthwhile to mention that in the construction of  EFT,  tensors are crucial. Let us give an example with two tensor operators $\hat{\cal O}_{A}$ and $\hat{\cal O}_{B}$ which can be defined as:
\bea \hat{\cal O}_{A}&=&\left(\hat{\cal O}^{(0)}_{A}+\delta \hat{\cal O}_{A}\right),~~~~ \hat{\cal O}_{B}=\left(\hat{\cal O}^{(0)}_{B}+\delta \hat{\cal O}_{B}\right).\eea
Here the superscript $(0)$ stands for tensor operators in unperturbed FLRW background and the rest of the part signify fluctuations on these operators. The construction rule of these operators give rise to a composite operator 
\bea \hat{\cal O}_{A}\hat{\cal O}_{B}&=&\left[-\hat{\cal O}^{(0)}_{A}\hat{\cal O}^{(0)}_{B}+\delta\hat{\cal O}_{A}\delta\hat{\cal O}_{B}+\hat{\cal O}^{(0)}_{A}\hat{\cal O}_{B}+\hat{\cal O}_{A}\hat{\cal O}^{(0)}_{B}\right].\eea
where
$\hat{\cal O}^{(0)}_{A}=\hat{\cal O}^{(0)}_{A}(g_{\mu\nu},n_{\mu},t)$ and $\hat{\cal O}^{(0)}_{B}=\hat{\cal O}^{(0)}_{B}(g_{\mu\nu},n_{\mu},t)$.
In our construction of EFT, possible choices for the tensor operator $\hat{\cal O}_{B}$ are extrinsic curvature $K_{\mu\nu}$ and Riemann tensor $R_{\mu\nu\alpha\beta}$. 

\item As mentioned earlier,  covariant derivative of the unit normal vector $n_{\mu}$ also plays a significant role in the construction of EFT. For  example:
\bea \int d^4x\sqrt{-g}J(t)K^{\mu}_{\mu}=\int d^4x\sqrt{-g}J(t)h_{\mu}^{\nu}\nabla_{\nu}n^{\mu}=\int d^4x \sqrt{-g}\sqrt{-g^{00}}\dot{J}(t).\eea
where $J(t)$ is a any arbitrary time dependent contribution appearing in the action.
\end{enumerate}

Using the above algorithm, a fairly general EFT action for inflationary cosmology can be written as
		\bea\label{ooo1}
		\mathcal{S}&=&\int d^4x \sqrt{-g}\left[\frac{1}{2}M_{p}^2R-\Lambda(t)-c(t)g^{00}+\delta{\cal O}_{\rm \bf EFT}\right]~~~~~~ 
		\eea

Since in this article we will be concentrating on two-point correlation functions only, both in the context of CMB and LSS, 
we will consider the following truncated EFT action derived using the EFT operators  \cite{crem}: 
   	 \bea \label{trunc}
   	     	\mathcal{S}&=&\int d^4x \sqrt{-g}\left[\frac{1}{2}M_{p}^2R-\left(\dot{H}+3H^2\right)M^2_p+\dot{H}M^2_pg^{00}+\frac{M^4_2(t)}{2}
   	     	(\delta g^{00})^2 \right.\nonumber\\ &&\left.~~~~~~~~~~~~~~~~~~
   	     	-\frac{\bar{M}^3_1(t)}{2}\delta g^{00}\delta K_{\mu}^{\mu}-\frac{\bar{M}^2_2(t)}{2}\left(\delta K_{\mu}^{\mu}\right)^2-\frac{\bar{M}^2_3(t)}{2}\delta K_{\mu}^{\nu} \delta K_{\nu}^{\mu}\right].~~~~~~ 
   	     	\eea
		
Note that this is a proper truncation as derived from the construction rules of EFT action, thereby leaving behind any ambiguity or
inconsistency in the development of the theory. Thus the above EFT action is a consistent one developed from
the first principles.

		
 \subsection{The role of Goldstone Boson}\label{2b}
  In this section, our prime objective is to construct a theory of Goldstone boson starting from the EFT proposed in the previous section. We will use the fact that the EFT operators and the corresponding action break temporal diffeomorphism explicitly. This is commonly known as {\it St$\ddot{u}$ckelberg mechanism} 
  \cite{stue1, stue2} in gravity where broken temporal diffeomorphism of Goldstone realizes the symmetry in non-linear way. This is  equivalent to the gauge transformation in $SU(N)$ non-abelian gauge theory where one studies the contribution from the longitudinal components of a massive gauge boson. In order  to establish broken temporal diffeomorphism we use the following transformations \bea t&\rightarrow& \tilde{t}=t+\xi^0(t,{\bf x});~~~~~
  x^{i}\rightarrow \tilde{x}^{i}=x^{i},\eea
  where $\xi^0(t,{\bf x})$ is a spacetime dependent parameter which is a measure of broken temporal diffeomorphism. See ref.~\cite{crem,sen1} for more details.
  
  To proceed further, we note that under these set of transformation equations the metric and its inverse components transform as~\footnote{It is important to note that, in the present context any general rank $m$ contravariant and covariant tensor transformation under broken temporal diffeomorphism as:
    \bea B^{\mu_{1}\mu_{2}\cdots\mu_{m}}&\rightarrow& \prod^{m}_{i=1}\left(\delta^{\mu_{i}}_{\alpha_{i}}+\delta^{\mu_{i}}_{0}\partial_{\alpha_{i}}\pi\right)B^{\alpha_{1}\alpha_{2}\cdots\alpha_{m}},\\  
     B_{\mu_{1}\mu_{2}\cdots\mu_{m}}&\rightarrow& \prod^{m}_{i=1}(\delta_{\mu_{i}}^{\alpha_{i}}+\delta^{\alpha_{i}}_{0}\partial_{\mu_{i}}\pi)^{-1}B_{\alpha_{1}\alpha_{2}\cdots\alpha_{m}}
    \eea }:
   \bea g^{\mu \nu}&\rightarrow& \tilde{g}^{\mu \nu}=\left(\frac{\partial \tilde{x}^\mu}{\partial x^{\alpha}}\right)\left(\frac{\partial \tilde{x}^\nu}{\partial x^{\beta}}\right)g^{\alpha\beta}=\left(\delta^{\mu}_{\alpha}+\delta^{\mu}_{0}\partial_{\alpha}\pi\right)\left(\delta^{\nu}_{\beta}+\delta^{\nu}_{0}\partial_{\beta}\pi\right)g^{\alpha \beta}
   ,\\
   g_{\alpha\beta}&\rightarrow&  \tilde{g}_{\alpha \beta}=\left(\frac{\partial x^\mu}{\partial \tilde{x}^{\alpha}}\right)\left(\frac{\partial x^\nu}{\partial \tilde{x}^{\beta}}\right)g_{\mu\nu} 
      \approx\left(\delta^{\mu}_{\alpha}-\delta^{\mu}_{0}\partial_{\alpha}\pi\right)\left(\delta^{\nu}_{\beta}-\delta^{\nu}_{0}\partial_{\beta}\pi\right)g_{\mu\nu},\eea
   which is perfectly consistent with the following constraint condition~\footnote{Once we use the following truncated expression for the series expansion:
   \be \left(\delta^{\mu}_{\alpha}+\delta^{\mu}_{0}\partial_{\alpha}\pi\right)^{-1}\approx \left(\delta^{\mu}_{\alpha}-\delta^{\mu}_{0}\partial_{\alpha}\pi\right),\ee
   we get one more additional constraint equation as given by:
   \be \left(\delta^{\alpha}_{\mu}+\delta^{\alpha}_{0}\partial_{\mu}\pi\right)\left(\delta^{\beta}_{\nu}+\delta^{\beta}_{0}\partial_{\nu}\pi\right)\left(\delta^{\sigma}_{\beta}-\delta^{\sigma}_{0}\partial_{\beta}\pi\right)\left(\delta^{\rho}_{\gamma}-\delta^{\rho}_{0}\partial_{\gamma}\pi\right)=\delta^{\alpha}_{\mu}\delta^{\sigma}_{\nu}\delta^{\rho}_{\gamma},\ee
   which is a necessary constraint to satisfy Eq~(\ref{df}).}:
   \bea \label{df} g^{\alpha\beta}g_{\beta\gamma}\rightarrow  \tilde{g}^{\alpha\beta}\tilde{g}_{\beta\gamma}=\delta^{\alpha}_{\mu}\delta^{\sigma}_{\nu}\delta^{\rho}_{\gamma}g^{\mu\nu}g_{\sigma\rho}=g^{\alpha\sigma}g_{\sigma\gamma}
   =\delta^{\alpha}_{\gamma}.\eea
   From these set of equations we get the following transformation rule  of the components of the contravariant metric: 
   \bea {g}^{00}\rightarrow\tilde{g}^{00}&=& 
   (1+\dot{\pi})^2 {g}^{00}+2(1+\dot{\pi}){g}^{0 i}\partial_{i}\pi+{g}^{ij}\partial_{i}\pi\partial_{j}\pi,~~~~~~~~\\ {g}^{0i}\rightarrow\tilde{g}^{0i}&=& 
      (1+\dot{\pi}){g}^{0i}+{g}^{ij}\partial_j \pi,\\
    {g}^{ij}\rightarrow\tilde{g}^{ij}&=&{g}^{ij},\eea
    and of the covariant metric:
    \bea g_{00}&\rightarrow& \tilde{g}_{00}= (1+\dot{\pi})^2 g_{00},\\
        g_{0i}&\rightarrow& \tilde{g}_{0i}= (1+\dot{\pi}){g}_{0i}+{g}_{00}\dot{\pi}\partial_{i}\pi,\\ g_{ij}&\rightarrow& \tilde{g}_{ij}= {g}_{ij}+{g}_{0j}\partial_{i}\pi+{g}_{i0}\partial_{j}\pi.\eea
    
  Having done that, let us now introduce a local field $\pi(x)\equiv \pi(t,{\bf x})$ identified to be the Goldstone mode that transforms under the broken temporal diffeomorphism as \cite{crem}:
    \be \pi(t,{\bf x})\rightarrow \tilde{\pi}(t,{\bf x})=\pi(t,{\bf x})-\xi^0(t,{\bf x}).\ee
   Since we work in unitary gauge for the calculations of cosmological perturbations,  the gauge fixing condition is given by
   $ \pi(x)=\pi(t,{\bf x})=0$,
  that results in
   $\tilde{\pi}(t,{\bf x})=-\xi^0(t,{\bf x})$.
It is worthwhile to note that, any time-dependent dynamical function under broken time diffeomorphism transform as \be f(t)\rightarrow f(t+\pi)=\left[\sum^{\infty}_{n=0}\frac{\pi^{n}}{n!}\frac{d^{n}}{dt^n}\right]f(t).\ee
    where the Taylor series expansion is terminated in appropriate order in Goldstone mode. Note that we can truncate this expansion for two-fold reasons:
        \begin{enumerate}
    \item During inflation Hubble parameter $H$ and its time derivative $\dot{H}$ does not change significantly so that quasi de Sitter approximation holds good. In the present computation under broken time diffeomorphism the Hubble parameter transform as:
    \bea\label{Ooiucc} H(t+\pi)&=&  \left[\sum^{\infty}_{n=0}\frac{\pi^{n}}{n!}\frac{d^{n}}{dt^n}\right]H(t),~~~~
            \eea
            which is the Taylor series expansion of the Hubble parameter by assuming the contribution from the Goldstone modes are small.
    Further we use, $\epsilon=-\dot{H}/H^2$ as the Hubble slow roll parameter in terms of which Eq~(\ref{Ooiucc}) can be recast as
    \bea\label{Ooiuccx} H(t+\pi)&=&\left[1-\pi H(t) \epsilon-\frac{\pi^2H(t)}{2}\left(\dot{\epsilon}-2\epsilon^2\right)+\cdots\right]H(t),~~~~
                \eea
    In the slow roll regime of inflation $\epsilon<<1$ and one can neglect all the higher order contributions in slow roll parameter $\epsilon$ and its time derivatives. 
    Following similar logic one can also  Taylor expand the term $c(t)$, which has been done explicitly in the next subsection.
    
    \item Let us consider the EFT Wilson coefficient $M_2$ which transform under broken time diffeomorphism as:
        \be\label{Ooiu} M_2(t+\pi)=  \left[\sum^{\infty}_{n=0}\frac{\pi^{n}}{n!}\frac{d^{n}}{dt^n}\right]M_2(t)
        \ee
         For further simplification we introduce here the canonically normalized Goldstone boson, which is defined as, $\pi_{c}=M_2^2 \pi$,
          using which Eq~(\ref{Ooiu}) can be recast as: 
            \bea \label{cvcv} M_2(t+\pi)&=& \left[\sum^{\infty}_{n=0}\frac{\pi^{n}_c}{n!M^{2n}_2}\frac{d^{n}}{dt^n}\right]M_2(t)\approx M_2(t).
            \eea
             It clearly implies that, if we go higher order in the Taylor series expansion then we will get additional suppression from $M_2$. For the other three coefficients, $\bar{M}_1$, $\bar{M}_2$ and $\bar{M}_3$ the higher order contributions in the Taylor series expansion get also suppressed by $M_2$ as appearing in Eq~(\ref{cvcv}).
    \end{enumerate} 
    Further, to construct the EFT action it is important to explicitly know the transformation rule of each an every contributions under broken time diffeomorphism, which are appended below:
    \begin{enumerate}
    \item The  3-hypersurface Ricci scalar and spatial component of the Ricci tensor transform as:
    \bea {}^{(3)}R&\rightarrow&{}^{(3)}\tilde{R}={}^{(3)}R+\frac{4}{a^2}H(\partial^2\pi),\\
    {}^{(3)}R_{ij}&\rightarrow&{}^{(3)}\tilde{R}_{ij}={}^{(3)}R_{ij}+H(\partial_{i}\partial_{j}\pi+\delta_{ij}\partial^2\pi),\eea
    where the operator $\partial^2=\partial^2_{k}$.
    \item The trace and spatial component of the extrinsic curvature tensor transform as:
        \bea \delta K&\rightarrow&\delta\tilde{K}=\delta K-3\pi\dot{H}-\frac{1}{a^2}(\partial^2\pi),\\
        \delta K_{ij}&\rightarrow&\delta\tilde{K}_{ij}=\delta K_{ij}-\pi\dot{H}h_{ij}-\partial_{i}\partial_{j}\pi.\eea
    \item The pure time component and mixed component of the extrinsic curvature tensor transform as:
    \bea \delta K^{0}_{0}&\rightarrow&\delta\tilde{K}^{0}_{0}=\delta K^{0}_{0},\\
    \delta K^{0}_{i}&\rightarrow&\delta\tilde{K}^{0}_{i}=\delta K^{0}_{i},\\
    \delta K^{i}_{0}&\rightarrow&\delta\tilde{K}^{i}_{0}=\delta K^{i}_{0}+2Hg^{ij}\partial_j\pi.\eea
    \end{enumerate}

   \subsection{Pathology of weak coupling approximation}\label{2c}
Let us now  discuss  the consequences of weak coupling approximation between gravity and Goldstone bosons on the EFT action Eq~(\ref{ooo1}). 
We will begin by writing down the transformation of the function $c(t)g^{00}$ of the action under  broken time diffeomorphism
  \bea\label{jkjk}  c(t)g^{00}=-\dot{H}M_{p}^2g^{00}\rightarrow c(t+\pi)\tilde{g}^{00}&=&-M_{p}^2\dot{H}(t+\pi)\left[ (1+\dot{\pi})^2g^{00}+2(1+\dot{\pi})\partial_i \pi g^{0i}~~~~~~~~~\right.\nonumber\\&&\left.~~~~~~~~~~~~~~~~~~~~~~~~~~~~~~~~~~~~~~~~~~~~~+g^{ij}\partial_i\pi \partial_j \pi\right],~~~~~~~~~\nonumber
  \\ 
  &=&c(t)\left[1+\frac{\pi}{\epsilon}\left(\dot{\epsilon}-2H\epsilon^2\right)+\cdots\right]\left[ (1+\dot{\pi})^2g^{00}~~~~~~~~~\right.\nonumber\\&&\left.~~~~~~~~~~~~~~~~~~~+2(1+\dot{\pi})\partial_i \pi g^{0i}+g^{ij}\partial_i\pi \partial_j \pi\right].\eea
  For further simplification we decompose the temporal component of the metric $g^{00}$ into background ($\bar{g}^{00}=-1$) and perturbations ( $\delta g^{00}$) as
  \bea \label{goo} g^{00}=\bar{g}^{00}+\delta g^{00},\eea
  substitute Eq~(\ref{goo}) in Eq~(\ref{jkjk}) and consider only the first term  appearing in the transformation rule.  It contains a Kinetic term, $M_{p}^2\dot{H}\dot{\pi}^2\bar{g^{00}}$ and a mixing term, $M_{p}^2\dot{H}\dot{\pi}\delta g^{00}$. From the mixing term we define another normalized field
   $\delta g^{00}_c=M_{p}\delta g^{00}$,
   in terms of which  the mixing term can be recast as
   \bea  M_{p}^2\dot{H}\dot{\pi}\delta g^{00}\rightarrow \sqrt{\dot{H}}\dot{\pi}_c\delta g^{00}_c.\eea 
   It is important to note that, this mixing term has one less derivative compared to the kinetic term. Consequently, in the energy limit, $ E>E_{mix}=\sqrt{\dot{H}}$, we can neglect this contribution from the EFT action. The mixing energy scale $E_{mix}$ is  the decoupling limit  above which one can completely  neglect the mixing between gravity and Goldstone boson fluctuation \cite{crem,sen1}.
   
 It may be mentioned here that  apart from the above mixing term, one can also have other  mixing contributions in the EFT action, which can be shown to be negligibly small.  Consider, for example, the term $M_{p}^2\dot{H}\dot{\pi}^2\delta{g^{00}}$. By recasting it after canonical normalization as
   \bea  M_{p}^2\dot{H}\dot{\pi}^2\delta{g^{00}}\rightarrow\frac{\dot{\pi}_c^2\delta{g^{00}_c}}{M_{p}}.\eea  it becomes obvious that this term is  Planck-suppressed. Similarly, just by doing dimensional analysis we can say that  further higher order contributions in $\dot{\pi}$ will lead to additional Planck-suppression in the EFT action. Consequently, we can safely consider  that $M_{p}^2\dot{H}\dot{\pi}\delta{g^{00}}$ is the leading order mixing term and hence neglect any contributions from the  mixing terms as long as we are working in the energy regime $E>E_{mix}$.

  In addition, from the expansion of $\dot{H}$, in the decoupling limit we get terms like, $\pi M_{p}^2\ddot{H}\dot{\pi}\bar{g}^{00}$, which can be recast as
    \bea \pi M_{p}^2\ddot{H}\dot{\pi}\bar{g}^{00}=-\pi\dot{\pi}M^2_pH^2\left(\dot{\epsilon}-2H\epsilon^2\right)\rightarrow\frac{\ddot{H}}{\dot{H}}\pi_c\dot{\pi}_c\bar{g}^{00}=\left(\frac{\dot{\epsilon}}{\epsilon}-2H\epsilon\right)\pi_c\dot{\pi}_c\bar{g}^{00}.\eea In the slow roll regime as, $\ddot{H}/\dot{H}<<1$, we can easily neglect this contribution from the EFT action. 
  This argument essentially implies that if we go higher in powers of $\pi$, then contributions are more suppressed by Planck scale after canonical normalization. 
  
 So, at the end of the day,  considering weak coupling approximation leads to  the following simplified expression for  the transformation rule of the function $c(t)g^{00}$  under broken time diffeomorphism in the decoupling limit
  \bea\label{jkjkx}  c(t)g^{00}=-\dot{H}M_{p}^2g^{00}\rightarrow c(t+\pi)\tilde{g}^{00}&\approx&c(t)g^{00}\left[\dot{\pi}^2-\frac{1}{a^2}(\partial_i\pi)^2\right].\eea
   In the above mentioned transformation rule we use the fact that liner $\pi$ terms are absent as the contributions from the tadpole diagrams are irrelevant in the EFT action \cite{sen1}.
   
   However, one needs to keep in mind that if one goes slightly below the decoupling scale, the weak coupling approximation still holds good and
   one can then consider the effect of the leading
   order mixing term  $M_{p}^2\dot{H}\dot{\pi}\delta g^{00}$ neglecting the subleading corrections. This will lead to relatively more accurate description of perturbations without violating any physical principle
   as such. We will elaborate on this in the next section.

 This completes the formal development of the properly truncated EFT action for inflationary cosmology.  Equipped with this, in the rest of the article we will engage ourselves in 
 analysing  quantum fluctuations during inflation and large scale structures therefrom. In the process we will confront our results with observations and search for 
 consistent constraints on the EFT parameters that lead to correct power spectrum both in CMB and LSS.

   \section{Quantum fluctuation from Goldstone modes}\label{sec-3}
  	\subsection{Two-point correlation function}\label{3a}
   To construct the two-point correlation function from the quantum fluctuation of the Goldstone modes we have to be consistent with derivatives of Goldstone.
   For our computation,
   we will   consider the contribution from the back reaction to be very small and  will also drop any contributions that contain quadratic derivatives of the Goldstone mode involving space and time.
   Consequently, one can drop all the other EFT parameters barring $\bar M_1$ and $M_2$ from the truncated action (\ref{trunc}).
   As a result,
   the second order  perturbed EFT action  for the decoupling regime turns out to be
   		\bea\label{eer}
   	\mathcal{S}^{(2)}
   	=\int dt~d^3x~ a^3\left(\frac{\frac{\bar{M_{1}}^3}{2}H-M_{p}^2\dot{H}}{c^2_S}\right)\left[\dot{\pi}^2-c^2_S\frac{(\partial_i \pi)^2}{a^2}\right],
   	\eea
   	where the effective sound speed squared parameter is defined as
   	\be \label{cs}
	c_S^2=\frac{\frac{\bar{M_{1}}^3}{2}H-M_{p}^2\dot{H}}{2M_2^4-M_{p}^2\dot{H}}.\ee
	See ref.~\cite{crem,zal,sen1} for the discussion on similar issues.
   	For further simplification one can express the spatial component of the metric fluctuation as
   	\bea\label{cc1} g_{ij}&=&a^2(t)\left[\left(1+2\zeta(t,{\bf x})\right)\delta_{ij}+\gamma_{ij}\right],\eea
   	where $\zeta(t,{\bf x})$  is the curvature perturbation that takes into account scalar fluctuations and the spin-$2$, transverse and traceless tensor $\gamma_{ij}$ represent tensor fluctuations. In order to exploit the correspondence between the Goldstone fluctuation and scalar fluctuation, we start with the transformation rule of the scale factor under broken time diffeomorphism, which can be expressed considering  terms upto linear order in Goldstone mode as
   	\bea\label{cc2} a(t)\rightarrow \tilde{a}(t-\pi)\approx a(t)\left[1-H\pi\right].\eea
   	Comparing Eq~(\ref{cc1}) and Eq~(\ref{cc2}) one readily obtains
   	$\zeta(t,{\bf x})=-H\pi(t,{\bf x})$ \cite{crem,sen1}
which shows explicitly the correspondence between curvature perturbation and Goldstone fluctuations.

   	Subsequently, the Mukhanov-Sasaki variable for this EFT setup can be defined  as \be v(\tau,{\bf x})=-z~\pi(\tau,{\bf x})HM_p,\ee where $z$ is defined as \be z=\left(\frac{\sqrt{2}\sqrt{\frac{\bar{M_{1}}^3}{2}H-M_{p}^2\dot{H}}}{c_S H M_p}\right)a=\frac{a}{c_S }\sqrt{\frac{\bar{M_{1}}^3}{HM^2_p}+2\epsilon}=2M_2^4-M_{p}^2\dot{H}.\ee
   	The scale factor for quasi de Sitter case can be expressed in terms of conformal time $\tau$ as
   	$a(\tau)=-\frac{1}{H\tau}(1+\epsilon)$.
   	In terms of the Mukhanov-Sasaki variable $v(\tau,{\bf x})$ the second order perturbed EFT action (\ref{eer}) can be recast as
   		\bea\label{eer111}
   	   	   	   	   	\mathcal{S}^{(2)}
   	   	   	   	   	=\frac{1}{2 }\int d\tau~d^3x~\left[{v}^{'2}+v^2\left(\frac{z^{'}}{z}\right)^2-2vv^{'}\left(\frac{z^{'}}{z}\right)-c^2_S(\partial_i v)^2\right],
   	   	   	   	   	\eea
   	   	   	   	   	where $(\cdots)^{'}\equiv d/d\tau$. Following usual technique with the  boundary conditions
   	   	   	   	   	\be \left[\frac{z^{'}}{z}\right]_{\del\Omega}=0, ~~~~
   	   	   	   	   	\left[v(\tau,{\bf x})\right]_{\del\Omega}=0,\ee
   	   	   	   	   	we arrive at  the following simplified version of the EFT action for scalar perturbation 
   	   	   	   	   	\bea\label{eer1111}
    	   	   	   	\mathcal{S}^{(2)}
   	   	  	   	   	=\frac{1}{2 }\int d\tau~d^3x~\left[{v}^{'2}+v^2\left(\frac{z^{''}}{z}\right)-c^2_S(\partial_i v)^2\right],
  	  	   	   	   	\eea
   	  	   	Further decomposition into Fourier modes of the Mukhanov-Sasaki variable $v(\tau,{\bf x})$ 
   	   	   	\bea v(\tau,{\bf x})&=&\int \frac{d^3k}{(2\pi)^3}v_{\bf k}(\tau)~e^{i{\bf k}.{\bf x}}\eea
   	   	   	leads to the expression for the perturbed EFT action for the scalar perturbation in Fourier space 
   \bea\label{eer11111}
      		   	\mathcal{S}^{(2)}
      		   	=\frac{1}{2 }\int d\tau~\frac{d^3k}{(2\pi)^3}~\left[{v^{'2}_{\bf k}(\tau)}+\left(k^2c^2_S-\frac{z^{''}}{z}\right)v^{2}_{\bf k}(\tau)\right],
      	   	   	   	   	   	   	   	   	   	\eea	
      	     	where $k=|{\bf k}|=\sqrt{{\bf k}.{\bf k}}$. Hence, as in the case of standard curvature perturbations, Eq~(\ref{eer11111}) represents an exactly parametric harmonic oscillator characterized by a time dependent frequency
      	   	\bea \omega^2(k,\tau)&=&\left(k^2c^2_S-\frac{z^{''}}{z}\right).\eea

Let us now pause for a moment and take  a careful look on 		
  the above EFT action (\ref{eer}) for scalar perturbations.
  The above action is strictly valid in the decoupling regime, i.e., at scales higher than the one at which gravity and Goldstone boson completely decouple. 
  As discussed in Section \ref{2c}, in this regime  one can completely drop all the terms arising from  mixing  between gravity and Goldstone Boson. So, strictly
  speaking,  this action will lead to results accurate upto order $\frac{H^2}{M_{p}^2}$ and $\epsilon$ \cite{bau, sen1}. However, 
  as discussed in \cite{bau, sen1}, the mass term of order $3 \epsilon H^2$ coming from the leading order mixing term $M_{p}^2\dot{H}\dot{\pi}\delta{g^{00}}$
  results in a more accurate expression for spectral tilt.
  As observational precision improve, 
  such as a 5-$\sigma$
detection of spectral index by Planck 2018 \cite{Aghanim:2018eyx,Akrami:2018odb}, our intention 
would be to obtain  results which are accurate at least upto next order leading to a better fit with observational
results. This can be materialized by modifying the action (\ref{eer}) by going slightly below the decoupling scale, where the weak coupling approximation still holds good.
		  One can then consider the effect of the leading
   order mixing term neglecting the subleading corrections. As mentioned in the previous section  \ref{2c}, this will lead to relatively more accurate description of perturbations without violating any physical principle
   as such.  

 Including the leading order mixing term leads to the following expression for the second order perturbed EFT action 
    
    		\bea\label{eer11}
    \mathcal{S}^{(2)}
    =\int dt~d^3x~ a^3\left(\frac{\frac{\bar{M_{1}}^3}{2}H-M_{p}^2\dot{H}}{c^2_S}\right)\left[\dot{\pi}^2-c^2_S\frac{(\partial_i \pi)^2}{a^2}+3\epsilon H^2 \pi^2\right],
    \eea
   
   A straightforward exercise as before leads to a parametric oscillator akin to Eq. (\ref{eer11111})  with modified time dependent frequency
    \bea \omega^2_{eff}(k,\tau)&=&\left(k^2c^2_S-\frac{z^{''}}{z}+3(aH)^2\epsilon \right).\eea
    that contains explicit dependence on the slow roll parameter $\epsilon$, and is hence more accurate.
    Consequently, 	the equation of motion in Fourier space can be written as
   	\be\label{fgfg} v_{\bf k}^{''}(\tau)+\omega^2_{eff}(k,\tau)v_{\bf k}(\tau)=0\ee
	
   In what follows	we will restrict our calculation upto first order in slow roll and first order in $\frac{2M_2^4}{M{pl}^2H^2\epsilon}$  for which
   	\be\begin{array}{lll}\label{qds} \displaystyle\frac{z''}{z}-3(aH)^2\epsilon \equiv\frac{1}{\tau^2}\left(\nu^2-\frac{1}{4}\right)\approx\frac{1}{\tau^2}\left[2+3\epsilon+\frac{3}{2}\eta+\frac{2M_2^4}{M_{p}^2H^2\epsilon}\left(-\frac{3\eta}{2}+\frac{3}{2}\epsilon\right)\right].~~~~~~\end{array}\ee
   For quasi de Sitter evolution of the background and considering the terms  in Eq~(\ref{qds}), the general solution to the equation of motion (\ref{fgfg}) can be readily obtained
   \be v_{\bf k}(\tau)=\sqrt{-\tau}\left[\alpha H_{\nu}^{(1)}(-c_{S}k\tau)+\beta H_{\nu}^{(2)}(-c_{S}k\tau)\right],\ee
   	where $\alpha$ and $\beta$ are the arbitrary integration constants and the numerical values of them are fixed by the choice of initial vacuum. For Bunch-Davies vacuum,   $\alpha=\sqrt{\pi/2}, ~
   	   \beta=0$ and
   	   the solution for the mode function finally boils down to
   	   \be v_{\bf k}(\tau)=\sqrt{-\frac{\pi\tau}{2}} H_{\nu}^{(1)}(-c_{S}k\tau).\ee
   	   where the argument for the Hankel function  $\nu$ is
	    \be \nu=\frac{3}{2}+\frac{1}{2}\left( 2\epsilon+\eta+\frac{2M_2^4}{M_{p}^2H^2\epsilon}\left[\epsilon-\eta\right]\right) \ee
	with  $\epsilon=-\frac{H'}{aH^2}$;~$\eta=\frac{\epsilon'}{aH\epsilon}$;~$\kappa=\frac{\eta'}{aH\eta}.$
	It   shows explicit  role of the EFT parameters and weak coupling pathology on the mode function that will further
	    reflect in two point function.

   To understand the behaviour of the solution let us consider two limiting cases $kc_{S}\tau\rightarrow -\infty$ and $kc_{S}\tau\rightarrow 0$, where the behaviour of the Hankel functions of the first kind are given by:
    \bea \lim_{kc_{S}\tau\rightarrow -\infty}H^{(1)}_{\nu}(-kc_S\tau)&=& \sqrt{\frac{2}{\pi}}\frac{1}{\sqrt{-kc_S\tau}}~e^{- ikc_S\tau}~e^{- \frac{i\pi}{2}\left(\nu+\frac{1}{2}\right)},\\
    \lim_{kc_{S}\tau\rightarrow 0}H^{(1)}_{\nu}(-kc_S\tau)&=& \frac{i}{\pi}~\Gamma(\nu)~\left(-\frac{kc_S\tau}{2}\right)^{-\nu}.\eea
   One can, in principle, analyse solutions for both super-horizon and sub-horizon modes. However, as is well-known, it is 
    the super-horizon modes that actually freeze out and appear as scalar perturbations  at a later epoch.
  Thus, they  have the major contribution to the two-point function.
  So, for all practical purpose, one can simply take into account the two-point function of  the super-horizon modes, which is given by
  \bea \langle \zeta(\tau,{\bf k})\zeta(\tau,{\bf q})\rangle&=&(2\pi)^3\delta^{(3)}({\bf k}+{\bf q})P_{\zeta}(k,\tau),\eea
  where $P_{\zeta}(k,\tau)$ is the primordial power spectrum for scalar fluctuation  at time $\tau$. Written explicitly in the context of
  EFT, it reads   \bea P_{\zeta}(k,\tau)&=&\frac{|v_{\bf k}(\tau)|^2}{z^2M^2_p}=\frac{2^{2\nu-3}H^2(-kc_S\tau)^{3-2\nu}}{4c_S(1+\epsilon)^2M^2_p.
   \left(\frac{\bar{M_{1}}^3}{HM^2_p}+2\epsilon\right)}\frac{1}{k^3}\left|\frac{\Gamma(\nu)}{\Gamma\left(\frac{3}{2}\right)}\right|^2~(1+k^2c^2_S\tau^2).~~~~~~~~\eea

  As a result, for the relevant modes for  CMB observables,
    the primordial power spectrum  at the horizon crossing $k_{*}c_S\tau=-1$ takes the following form

       \bea \label{two-point}
        P_{\zeta}(k_*)&=&\frac{(-c_s\tau)^3 2^{2\nu-3}H^2}{4c_S(1+\epsilon)^2M^2_p.
         \left(\frac{\bar{M_{1}}^3}{HM^2_p}+2\epsilon\right)}\left|\frac{\Gamma(\nu)}{\Gamma\left(\frac{3}{2}\right)}\right|^2=2^{2\nu}\frac{(-\tau)}{4\pi} (\Gamma(\nu))^2\frac{H^2}{a^2(-M_{p}^2\dot{H}+2M_2^4)}.~~~~~~~~~~
         \eea
       As usual, one can also define a dimensionless power spectrum therefrom
      $\Delta_{\zeta}(k)=\Delta_{\zeta}(k_*)=\frac{1}{2\pi^2}P_{\zeta}(k_*)$,
      where $k_*$ is the pivot scale on which $P_{\zeta}(k_*)$ and $\Delta_{\zeta}(k_*)$ are evaluated.
         
Consequently, the spectral tilt turns out to be
                       \bea \label{errrr} n_{\zeta}(k_*)-1&=&\left[\frac{d\ln\Delta_{\zeta}(k)}{d\ln k}\right]_{k_{*}c_S\tau=-1}\nonumber\\
           &=& (3-2\nu	)\nonumber\\&=& -\left(  2\epsilon+\eta+\frac{2M_2^4}{M_{p}^2H^2 \epsilon}\left[\epsilon-\eta\right] \right) \eea
   



     \section{Structures from EFT}\label{sec-4}
     
The regime of LSS can be an important probe for primordial cosmology. Modes of perturbations that 
exits during inflation are the seeds of Large Scale Structure we 
see today. As the the EFT
parameters affect the primordial powerspectrum \eqref{two-point}, these two parameters will also 
affect the growth of LSS. As mentioned earlier LSS surveys have much larger range of probable scales
than any primordial survey, LSS survey can constrain the primordial parameters with good accuracy, so 
probing the EFT parameters $M_2$ and $\bar{M}_1$ that appear in \eqref{two-point} with LSS
surveys will be relevant and interesting. First let us briefly review the linear and non-linear regimes
of LSS.


   \subsection{Linear and non-linear regimes: transfer function}\label{4a}
   
 
At large scales,  i.e., in the linear regime, the power spectrum is given by the primordial power spectrum with slight spectral tilt as derived earlier. 
At relatively smaller scales, where the non-linear regime sets in, the behaviour of perturbations can be manifestly written in the language of a 
 momentum dependent fitting function, called transfer function $T(k)$, that basically transfers the primordial power to smaller scales.
 This can be expressed in terms of the gravitational potential as
    \bea \label{ddf} \Phi(a,{\bf k})&=&\frac{3}{2}\frac{H^2_0}{ak^2}\Omega_{m}\delta_{g}({a,{\bf k}})=\frac{9}{10}\Phi_{\rm prim}({\bf k})T(k)\frac{D_{g}(a)}{a},\eea
   where $\Phi_{\rm prim}({\bf k})$ is the primordial potential and $D_{g}(a)$ is the  growth function. Written explicitly, 
      \bea D_{g}(a)&=&\frac{5}{2}\Omega_{m}\frac{H(a)}{H_0}\int^{a}_{0}\left(\frac{H_0}{\tilde{a}H(\tilde{a})}\right)^3d\tilde{a}.\eea
      For flat, matter dominated universe it simply boils down to $a$.

The matter overdensities can then be conveniently written as 
    \bea \label{ddf1} \delta_{g}({a,{\bf k}})=\frac{2}{3}\Phi(a,{\bf k})\frac{ak^2}{\Omega_m H^2_0}=-\frac{2}{3}\zeta(\tau,{\bf k})\frac{ak^2}{\Omega_m H^2_0}=\frac{3}{5}\Phi_{\rm prim}({\bf k})T(k)D_{g}(a)\frac{k^2}{\Omega_m H^2_0}.\eea


   An age-old  transfer function is given by \cite{weinbook}
 \be\begin{array}{lll}
  \displaystyle T(k)\displaystyle=\left\{\begin{array}{ll}
                     \displaystyle  1 &
  \mbox{\small { for $k_{IR}<k<k_{eq}$}}  
 \\ 
          \displaystyle  12\left(\frac{k_{eq}}{k}\right)^2 \ln\left(\frac{k}{8k_{eq}}\right),~~~~~~~~~~ & \mbox{\small { for $k_{eq}<k<k_{UV}$}}.
           \end{array}
 \right.
 \end{array}\ee
    where $k_{eq}$ is the corresponding momentum scale for matter-radiation equality which can be expressed in terms of matter abundance ($\Omega_mh^2$) as: \be k_{eq}=a_{eq}H(a_{eq})=0.073~{\rm Mpc}^{-1}~\Omega_mh^2.\ee
    
 However,  the above transfer function is not too useful as  it has a discontinuity near $k_{eq}$. Further, given latest data, one needs to have a more accurate
 fitting function for the same that can describe the observable universe more accurately. Couple of good fitting functions are available in the literature
 that can more or less successfully serve the purpose like BBKS \cite{bbks} and Eisenstein-Hu \cite{eisensteinhu}.
 
\subsection{Loop correction}

 Loop corrections of the powerspectrum is important to understand the mode-mixing in the non-linear
 regime. With the tree level matter powerspectrum that can be calculated using \eqref{two-point},
 \eqref{ddf} and the choice of transfer function we can calculate the higher order corrections.
   Considering the contributions upto one-loop in the non-linear perturbation we get the following simplified expansion for the total two-point function :
         \bea \langle \delta_{g}({\tau,{\bf k}})\delta_{g}({\tau,{\bf q}})\rangle&=&\sum^{1}_{n=0}\langle \delta_{g}({\tau,{\bf k}})\delta_{g}({\tau,{\bf q}})\rangle^{(n)}\nonumber\\
         &=&\left[\langle \delta_{g}({\tau,{\bf k}})\delta_{g}({\tau,{\bf q}})\rangle^{(0)}+\langle \delta_{g}({\tau,{\bf k}})\delta_{g}({\tau,{\bf q}})\rangle^{(1)}+ \cdots \right],~~~~~~~~~~\eea
         where the individual contributions are given by
        \bea \label{ww1}\langle \delta_{g}({\tau,{\bf k}})\delta_{g}({\tau,{\bf q}})\rangle^{(0)}&=&\langle \delta_{g}({\tau,{\bf k}})\delta_{g}({\tau,{\bf q}})\rangle_{\bf Tree}\nonumber\\&=&\langle \delta^{(1)}_{g}({\tau,{\bf k}})\delta^{(1)}_{g}({\tau,{\bf q}})\rangle,\\
        \label{ww2}\langle \delta_{g}({\tau,{\bf k}})\delta_{g}({\tau,{\bf q}})\rangle^{(1)}&=&\langle\delta_{g}({\tau,{\bf k}})\delta_{g}({\tau,{\bf q}})\rangle_{\bf 1-loop}\nonumber\\
        &=&\langle\delta^{(2)}_{g}({\tau,{\bf k}})\delta^{(2)}_{g}({\tau,{\bf q}})\rangle+2\langle\delta^{(1)}_{g}({\tau,{\bf k}})\delta^{(3)}_{g}({\tau,{\bf q}})\rangle
        ,\\
          \cdots \cdots \cdots&&\cdots \cdots \cdots\cdots \cdots \cdots. \nonumber
         \eea
         Here in Eq~(\ref{ww2}) a factor of $2$ is appearing due to the symmetry in the cross correlations
          \bea \langle\delta^{(i)}_{g}({\tau,{\bf k}})\delta^{(j)}_{g}({\tau,{\bf q}})\rangle&=&
          \langle\delta^{(j)}_{g}({\tau,{\bf k}})\delta^{(i)}_{g}({\tau,{\bf q}})\rangle~~~~\forall i,j=1,2,\cdots{\rm with}~~i\neq j.\eea
        Consequently, the total matter power spectrum for the overdensity field 
    \be \label{kio}
    P_{\delta_{g}}(k,\tau)=\sum^{\infty}_{n=0}P^{(n)}_{\delta_{g}}(k,\tau)
     =\sum^{\infty}_{i,j=1,i+j={\rm \bf even}}P^{(ij)}_{\delta_{g}}(k,\tau).
    \ee
 reduces to the following truncated power spectrum  upto one-loop correction:
    \bea \label{po1} 
    P_{\delta_{g}}(k,\tau)&=&\sum^{1}_{n=0}P^{(n)}_{\delta_{g}}(k,\tau)=\left[P^{(0)}_{\delta_{g}}(k,\tau)+P^{(1)}_{\delta_{g}}(k,\tau)\right].\eea
  Finally,  comparing Eq~(\ref{kio}) and Eq~(\ref{po1}), and collecting the contributions from 
    tree level and one loop separately, one arrives at
    \bea P^{(0)}_{\delta_{g}}(k,\tau)&=&P^{\bf Tree}_{\delta_{g}}(k,\tau)=P^{(11)}_{\delta_{g}}(k,\tau),\\
    P^{(1)}_{\delta_{g}}(k,\tau)&=&P^{\bf 1-loop}_{\delta_{g}}(k,\tau)=P^{(22)}_{\delta_{g}}(k,\tau)+2P^{(13)}_{\delta_{g}}(k,\tau)\eea

Though these analytical computations help visualize the complicated mode-mixing, it is  difficult
to compute this loop corrections analytically with the powerspectrum of \eqref{two-point}. To use the LSS 
regime for estimations of these EFT parameters we will apply Fisher forecast technique using
upcoming surveys in the next section.

        
\section{EFT Parameters and Fisher Forecast}\label{sec-5}

Fisher matrix forecast analysis allows us to estimate 
expected measurement errors when marginalizing over a set of 
parameters. By calculating the Fisher information matrix (the inverse of co-variance matrix) the 
error bars can be estimated for upcoming surveys \cite{Tegmark:1996bz,Wolz:2012sr}.
In this section we would use Fisher forecast method to estimate the EFT parameters $M_2$ and $\bar{M}_1$  for LSST survey \cite{Abell:2009aa} and EUCLID survey \cite{Laureijs:2011gra}. 
The LSST and EUCLID surveys are two of the most sensitive upcoming LSS surveys which can probe the
inflationary dynamics with high precision.
 The BAO probe 
of LSST is sensitive to matter powerspectrum with 2.6 billion galaxies in the LSST 
`Gold Sample' \cite{Zhan:2017uwu}. EUCLID survey will also probe the LSS regime with wide range of
probable scale and it can at least measure the spectral index $n_s$ with similar precision
 to Planck \cite{Amendola:2016saw}.

We have used the publicly available code COOP \cite{Huang:2015srv} for forecasting on the parameters.
COOP can calculate Fisher information matrix for the model parameters and 
put constraint on the parameters   
corresponding to the specified survey. One can modify the primordial powerspectrum in the code and 
implementation of powerspectrum 
\eqref{two-point} in COOP gives the following vector of parameters,

\begin{equation}
\lbrace H, \epsilon, \eta, M_2, \bar{M}_1\rbrace.
\end{equation}

The fiducial values of the parameters are assigned according to the constraints on them from
latest Planck data \cite{Akrami:2018odb}. The latest constraint from Planck 2018 on
 $H$, $\epsilon$ and $\eta$ are summarised in Table \ref{planck-bound}.
 As discussed in \ref{3a} we have calculated the primordial
observables upto first order in $\frac{2 M_2^4}{M_{pl}^2 H^2 \epsilon}$, 
 we have used the theoretical constraint on parameter 
$M_2$ as $M_2^4 \leq M_{pl}^2 \dot{H}$, to assign the fiducial value. We have explored two situations,
firstly parameter $\bar{M}_1$ is of the same order as $M_2$ and secondly parameter $M_2$ larger than
$\bar{M}_1$ to  make our analysis complete. The fiducial value of $\bar{M}_1$ is assigned according
to the case we have studied. In all the cases we have considered $M_2, \bar{M}_1 >0$.

\begin{table}
\begin{center}
	\begin{tabular}{ c c }
		\hline
		Parameter & Planck Constraint  \\
		\hline
		$H$ & $< 2.5 \times 10^{-5} ~~(95 \% \text{C.L.})$  \\
		$\epsilon$ & $<0.0063~~~~~~~(95 \% \text{C.L.})$ \\
		$\eta$ & $0.032^{+0.009}_{-0.008}~~~~ (68 \% \text{C.L.})$\\
		\hline
	\end{tabular} 
\caption{Planck constraints on the relevant parameters. $H$ is expressed in units of $M_{pl}$.}
\label{planck-bound}	
\end{center}
\end{table}

\subsection{Case I : $ M_2 \sim \bar{M}_1$}
First we consider parameter $\bar{M}_1$  to be of the same 
order of $M_2$ and same fiducial value is assigned to both of the parameters.

\begin{table}[]
\begin{center}
\begin{tabular}{|c|c|c|c|}
\hline
\multirow{2}{*}{Parameter} & \multirow{2}{*}{Fiducial Value} & \multicolumn{2}{c|}{Standard Deviation} \\ \cline{3-4} 
                           &  & LSST  & EUCLID             \\
\hline                           
                      H  & $1.5 \times 10^{-5}$ & $0.22e-5$ &  $0.73e-5$                  \\
                $\epsilon$ & $0.005$ & $0.0015$   &   0.0032                 \\
                 $\eta$ & $0.032$ & 0.11  & 0.25                    \\
                 $M_2$ & $0.001$ & $0.40e-4$ & $0.10e-3$                  \\
                 $\bar{M}_1$ & $0.001$ & $0.0058$ & $0.0080$\\
\hline
\end{tabular}
\caption{Fiducial value and standard deviation of the parameters for LSST and EUCLID survey with
$M_2=\bar{M}_1=0.001$. Parameter $H$, $M_2$, $\bar{M}_1$ are expressed in units of $M_{pl}$.}
\label{std-1}
\end{center}
\end{table}



The fiducial values and standard deviation of the parameters are summarized in 
Table \ref{std-1}. The standard deviation of $M_2$ is small with the fiducial value assigned but
the standard deviation on the second slow-roll parameter $\eta$ is large.
Also the standard deviation in $\bar{M}_1$ is too large
for both of the surveys. From \eqref{cs} we know that the sound speed is a function of $M_2$
and $\bar{M}_1$ but the large standard deviation of $\bar{M}_1$ leaves $c_s$ unconstrained.

\begin{table}[]
\begin{center}
\begin{tabular}{|c|c|c|c|}
\hline
\multirow{2}{*}{Parameter} & \multirow{2}{*}{Fiducial Value} & \multicolumn{2}{c|}{Standard Deviation} \\ \cline{3-4} 
                           &  & LSST  & EUCLID             \\
\hline                           
                      H  & $1.5 \times 10^{-5}$ & $0.98e-3$ &  $0.21e-2$           \\
                $\epsilon$ & $0.005$ & $0.074$   &   $0.14$          \\
                 $\eta$ & $0.032$ & $0.43$  & $0.78$           \\
                 $M_2$ & $0.0007$ & $0.00073$ & $0.0016$           \\
                 $\bar{M}_1$ & $0.0007$ & $5.22$ & $11.32$\\
\hline
\end{tabular}
\caption{Fiducial value and standard deviation of the parameters for LSST and EUCLID survey with
$M_2=\bar{M}_1=0.0007$. Parameter $H$, $M_2$, $\bar{M}_1$ are expressed in units of $M_{pl}$.}
\label{std-2}
\end{center}
\end{table}


 To show the dependence on the fiducial value we change the fiducial value of $M_2$ and $\bar{M}_1$
 to $0.0007 M_{pl}$ keeping fiducial values of other parameters same. From \ref{std-2} we can see that changing the fiducial value of $M_2$ and $\bar{M}_1$
 to $0.0007 M_{pl}$ the standard deviation of all the parameters worsen.
 
\subsection{Case II :  $M_2 > \bar{M}_1$} 
  
  Next we consider a scenario where parameter $M_2$ is larger than $\bar{M}_1$.


\begin{table}[]
\begin{center}
\begin{tabular}{|c|c|c|c|}
\hline
\multirow{2}{*}{Parameter} & \multirow{2}{*}{Fiducial Value} & \multicolumn{2}{c|}{Standard Deviation} \\ \cline{3-4} 
                           &  & LSST  & EUCLID             \\
\hline                           
                      H  & $1.5 \times 10^{-5}$ & $0.20E-5$ &  $0.67e-5$           \\
                $\epsilon$ & $0.005$ & $0.0014$   &   $0.0030$          \\
                 $\eta$ & $0.032$ & $0.11$  & $0.24$           \\
                 $M_2$ & $0.001$ & $0.40e-4$ & $0.10e-3$           \\
                 $\bar{M}_1$ & $0.0001$ & $0.51$ & $0.68$\\
\hline
\end{tabular}
\caption{Fiducial value and standard deviation of the parameters for LSST and EUCLID survey with
$M_2=0.001$, $\bar{M}_1=0.0001$. Parameter $H$, $M_2$, $\bar{M}_1$ are expressed in units of $M_{pl}$.}
\label{std-3}
\end{center}
\end{table} 


\begin{table}[]
\begin{center}
\begin{tabular}{|c|c|c|c|}
\hline
\multirow{2}{*}{Parameter} & \multirow{2}{*}{Fiducial Value} & \multicolumn{2}{c|}{Standard Deviation} \\ \cline{3-4} 
                           &  & LSST  & EUCLID             \\
\hline                           
                      H  & $1.5 \times 10^{-5}$ & $0.41e-4$ &  $0.59e-4$           \\
                $\epsilon$ & $0.005$ & $0.011$   &   $0.019$          \\
                 $\eta$ & $0.032$ & $0.42$  & $0.71$           \\
                 $M_2$ & $0.0007$ & $0.76e-4$ & $0.13e-3$           \\
                 $\bar{M}_1$ & $0.7e-4$ & $13.99$ & $13.99$\\
\hline
\end{tabular}
\caption{Fiducial value and standard deviation of the parameters for LSST and EUCLID survey with
$M_2=0.0007$, $\bar{M}_1=0.00007$. Parameter $H$, $M_2$, $\bar{M}_1$ are expressed in units of $M_{pl}$.}
\label{std-4}
\end{center}
\end{table}


From Table \ref{std-3} we can again see that with the standard deviation for parameter $M_2$ is small
for both the surveys and standard deviation for the parameter for $\bar{M}_1$ is large. The standard
deviation of $\eta$ is large like the previous cases.
Choosing
another set of fiducial value $M_2=0.0007 M_{pl}$ and $\bar{M}_1=0.00007 M_{pl}$, 
from Table \ref{std-4} we can see 
that the standard deviation of all the parameters again worsen with respect to Table \ref{std-3}. It is interesting to see that the standard 
deviation for $M_2$ implies both the surveys 
suggest a non-zero value for the parameter. This is different than the case of $M_2 \sim \bar{M}_1$
with fiducial value $M_2 = \bar{M}_1 = 0.0007 M_{pl}$ where $M_2$ had a larger
standard deviation. Also comparing Table \ref{std-1} and \ref{std-3} where only difference lies in the 
fiducial value of $\bar{M}_1$ it can be seen that the standard deviation of all the parameters 
remains the same except for parameter $\bar{M}_1$. So the surveys are not sensitive enough on 
$\bar{M}_1$.

\subsection{Comparison and major outcome of the analysis}

The key findings of the present analysis for the two different cases as done above can be summarized by the following points:

\begin{itemize}
\item The estimated error bars on $M_2$ are comparatively small considering the fiducial value $M_2 = 0.001 M_{pl}$ for both the cases. For the fiducial value of $M_2 = 0.0007 M_{pl}$ the errors get worsen, but $M_2 > \bar{M}_1$ case gives better error estimate.
\item In each case both the LSST and EUCLID survey are insensitive to the parameter $\bar{M}_2$.
\item As the parameter $\bar{M}_2$ is not well constrained the sound speed $c_s$ is not constrained.
\item The estimated error on second slow roll parameter $\eta$ gets worsen compared to Planck observation. This is expected as $c_s$, $\epsilon$, $\eta$, $M_2$ and $\bar{M}_1$ can have degeneracies.
\item In each of the case it is noticeable that LSST survey gives better error estimate than EUCLID survey.
\end{itemize}

\section{Summary}

In this article we have tried to probe the physics of inflation with LSS surveys. Starting from the 
framework of Effective Field Theory of inflation  we have studied quantum fluctuation for Goldstone bosons and computed the two-point correlation function for primordial  scalar perturbations.
With proper truncation of EFT Lagrangian we have found that the powerspectrum of scalar perturbations
are described by couple of EFT parameters and sound speed of scalar perturbation along with 
spectral tilt  also get modified by these parameters. 

These two EFT parameters also affect the physics of LSS, so the LSS surveys have the potential to
constrain these EFT parameters which will help us understand the proper dynamics of inflationary 
cosmology. To these end we have applied the Fisher forecast method
to estimate the error bars on these parameters for upcoming LSS surveys LSST and EUCLID. We have 
considered two cases $M_2 \sim \bar{M}_1$ and $M_2 > \bar{M}_1$. In the first case where 
$M_2 \sim \bar{M}_1$ we have found that the standard deviation for the parameter $M_2$ is small
for fiducial value $0.001 M_{pl}$ for both the surveys, but standard deviation for $\bar{M}_1$ is
large. This leaves the sound speed of perturbation unconstrained. Changing the fiducial value for 
both $M_2$ and $\bar{M}_1$ to $0.0007 M_{pl}$ we have noticed that the standard deviation of every 
parameters get large. Next we have considered $M_2 > \bar{M}_1$ with fiducial value 
$M_2 = 0.001 M_{pl}$ and $\bar{M}_1 = 0.0001 M_{pl}$ the results are the same as first case with
fiducial value $M_2 = \bar{M}_1 = 0.001 M_{pl}$, except the standard deviation of $\bar{M}_1$ is larger.
Changing the fiducial value to $M_2 = 0.0007 M_{pl}$ and $\bar{M}_1 = 0.00007 M_{pl}$, the 
standard deviation of the parameters gets larger. Comparing these two scenarios it can be seen that
both of LSST and EUCLID are not sensitive to the parameter $\bar{M}_1$ also in all the cases 
the standard deviation of second slow roll parameter $\eta$ is large.

To wind it up, the description of EFT of inflation encodes different model information in the EFT parameters. 
Constraining these parameters with data can give us a better understanding of the model of inflation.
Fisher matrix forecast method can give us an idea about the expected error bars on the parameters for the 
upcoming surveys, comparing our scenario with real data when the surveys become operational will be
very interesting.
We believe the analysis would  help the community to have a better understanding of the primordial physics as well as the EFT of inflation in the light of upcoming LSS surveys
once they are operational.

	\section*{Acknowledgments}
	
We gratefully acknowledge  computational facilities of ISI Kolkata.	
The research fellowship of SC is supported by the J.  C.  Bose National Fellowship of Sudhakar Panda.  SC also would line to thank School of Physical Sciences, National Institute for Science Education and Research (NISER), Bhubaneswar for providing the work friendly environment.  
AB would like to thank the computational facilities of ISI Kolkata and the University of Minnesota, Twin Cities where part of this work was done.
AB also thanks ISI Kolkata for supporting a long term project. 
SP thanks Department of Science and Technology, Govt. of India for partial support through Grant no. NMICPS/006/MD/2020-21.

   

\end{document}